\documentclass[aps,prl,showpacs,twocolumn,superscriptaddress,
	      floats,floatfix,showkeys]{revtex4}
\usepackage{amssymb,amsmath}
\usepackage{bm,url}
\usepackage{graphics}
\usepackage{subfigure}
\usepackage{epsfig}
\usepackage{ulem}
\usepackage{color}
\graphicspath{{./figures/}}
\def \kinj {k_{\rm inj}}

\def \bfu {{\bf u}}

\def \bfu {{\bf u}}

\def \curl {{\bm \nabla} \times}

\def \lap {\nabla^2}

\newcommand{\bra}[1]{\langle #1\rangle}

\def \Ttran  {T_{\rm tran}}
\def \Teddy {T_{\rm eddy}}
\def \TpE {T^+_{\rm E}}

\def \TmE {T^-_{\rm E}}
\def \TmL {T^-_{\rm L}}

\def \Tinj {T_{\rm inj}}
\def \Tcutoff {T_{\rm cf}}
\def \Teta {T_{\eta}}

\def \Tmean {T^+_{\rm mean}}

\def \urms  {u_{\rm rms}}

\def \np  {N_{\rm p}}

\def \ppe  {P^+_{\rm E}}
\def \ppl  {P^+_{\rm L}}
\def \pne  {P^-_{\rm E}}
\def \pnl  {P^-_{\rm L}}
\def \qpe  {Q^+_{\rm E}}
\def \qpl  {Q^+_{\rm L}}
\def \qne  {Q^-_{\rm E}}
\def \qnl  {Q^-_{\rm L}}

\urldef\ytubeurl\url{http://www.youtube.com/watch?v=ANDUNDHYgzk }{}

\definecolor{darkgreen}{rgb}{0,0.625,0}

\begin{document}
\title{Persistence Problem in Two-Dimensional Fluid Turbulence}
\author{Prasad Perlekar}
\email{p.perlekar@tue.nl}
\affiliation{ Department of Physics and Department of Mathematics and 
Computer Science. Eindhoven University of Technology, P.O. Box 513, 
5600 MB Eindhoven, The Netherlands} 
\author{Samriddhi Sankar Ray} 
\email{samriddhisankarray@gmail.com}
\affiliation{Centre for Condensed Matter Theory, Department of Physics, Indian
Institute of Science, Bangalore 560012, India.}
\affiliation{Laboratoire Cassiop\'ee, Observatoire de la C\^ote 
d'Azur, UNS, CNRS, BP 4229, 06304 Nice Cedex 4, France.}
\author{Dhrubaditya Mitra} 
\email{dhruba.mitra@gmail.com}
\affiliation{NORDITA, Roslagstullsbacken 23, SE-10691 Stockholm, Sweden }
\author{Rahul Pandit}
\email{rahul@physics.iisc.ernet.in}
\altaffiliation[\\ also at~]{Jawaharlal Nehru Centre For Advanced
Scientific Research, Jakkur, Bangalore, India.}
\affiliation{Centre for Condensed Matter Theory, Department of Physics, Indian
Institute of Science, Bangalore 560012, India.} 
\begin{abstract}
We present a natural framework for studying the persistence problem in
two-dimensional fluid turbulence by using the Okubo-Weiss parameter $\Lambda$
to distinguish between vortical and extensional regions. We then use a direct
numerical simulation (DNS) of the two-dimensional, incompressible
Navier--Stokes equation with Ekman friction to study probability distribution
functions (PDFs) of the persistence times of vortical and extensional regions
by employing both Eulerian and Lagrangian measurements. We find that, in the
Eulerian case, the persistence-time PDFs have exponential tails; by contrast,
this PDF for Lagrangian particles, in vortical regions, has a power-law tail
with an exponent $\theta=2.9\pm0.2$.
\end{abstract}
\keywords{turbulence, statistical mechanics, persistence}
\pacs{47.27.-i,05.40.-a }
\maketitle

The persistence problem, which is of central importance in nonequilibrium
statistical mechanics~\cite{maj99}, is defined as follows: For a fluctuating
field $\phi$, the persistence-time probability distribution function (PDF)
$P^\phi(\tau)$ yields the probability that the sign of $\phi$ at a point in
space does not change up to a time $\tau$. Theoretical, experimental, and
numerical studies of a variety of systems, ranging from reaction-diffusion
systems to granular media, have shown that $P^\phi(\tau) \sim \tau^{-\theta}$
as $\tau \rightarrow \infty$, where $\theta$ is the {\it persistence
exponent}. This nontrivial exponent 
cannot be obtained from dimensional arguments; it
can be calculated analytically only for a few
models~\cite{maj96+der96}; for most models  it
has to be obtained numerically. 
We propose a natural way of defining the
persistence problem in two-dimensional turbulence. We then show how to obtain
the persistence exponent for this case.  

Turbulent flows in two-dimensional fluid films display vortical points and
strain-dominated or extensional points. We show how to examine the {\it
persistence} of such points in time by direct numerical simulations (DNSs) of
the forced, two-dimensional, incompressible Navier--Stokes equation. Our
study has been designed with thin-fluid-film experiments in
mind~\cite{riv01,per+pan09} so we account for an air-drag-induced Ekman
friction and we drive the fluid by using a Kolmogorov forcing. We demonstrate
that the Okubo-Weiss parameter~\cite{oku70+wei91,per+cho87} $\Lambda$, whose
sign at a given point determines whether the flow there is vortical or
extensional, provides us with a natural way of studying such persistence.

It is important to distinguish the following types of persistence times: (A)
In the Eulerian framework we consider a point $(x,y)$ and, by following the
time evolution of $\Lambda$, determine the time $\tau$ for which the flow at
this point remains vortical~(extensional) if the flow at this point became
vortical~(extensional) at some earlier time; (B) in the Lagrangian framework
we consider how long a Lagrangian particle resides in a
vortical~(extensional) region if this particle entered that
vortical~(extensional) region at an earlier time. For all these cases we
obtain PDFs of the persistence or
residence times that we denote generically by $\tau$. We find, in the
Eulerian framework, that the PDFs of $\tau$ show exponential tails in both
vortical and extensional regions.  In the Lagrangian framework the PDF of the
residence time of the particle in extensional regions also shows an
exponential tail; the analogous PDF for vortical regions shows {\it a
power-law tail}. The persistence exponent that characterizes this power law
is independent of parameters such as the Reynolds number, the characteristic
scale of the forcing, and the coefficient of Ekman friction, at least at the
level of our numerical studies.

\begin{figure}
\begin{center}
\includegraphics[width=0.45\linewidth]{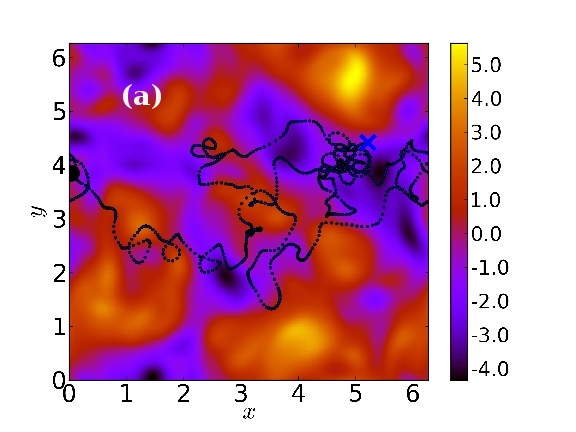}
\includegraphics[width=0.45\linewidth]{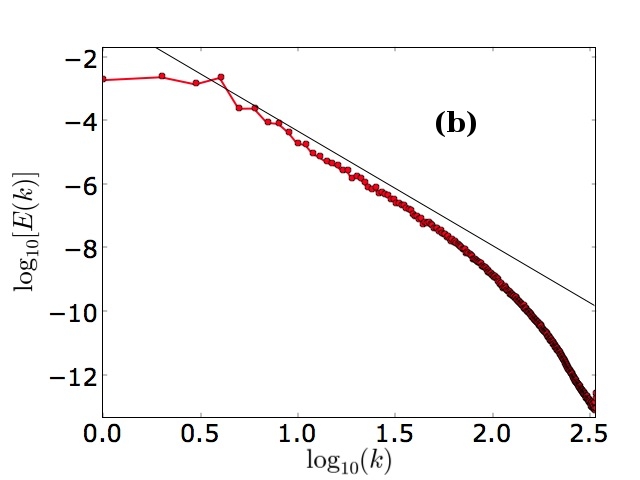}
\caption{\label{fig:snap} (Color online) (a) A pseudocolor
  plot of the stream function $\psi$, at a representative time in the
  statistically steady state, with a representative Lagrangian
  particle track (blue squares) superimposed on it from our run~${\tt
    R2}$. The symbol ${\rm o}$ indicates the beginning of the
  trajectory and the $\times$ sign marks its end. For an animated
  version see the movie file at \ytubeurl{}\/; (b) log-log 
  (base 10)  plot
  of the energy spectrum for our run ${\tt R4}$
  (\textcolor{red}{line with dots}); the black line with a slope 
  $-3.6$ is shown for reference.}
\end{center}
\end{figure}

We perform a direct numerical simulation (DNS) of the incompressible,
two-dimensional, Navier-Stokes equation 
\begin{equation} 
\partial_t \omega - J(\psi,\omega) = \nu \nabla^2 \omega + f_{\omega}  
- \mu \omega , 
\label{eqch6:nsvor1} 
\end{equation} 
with periodic boundary conditions, by using a pseudospectral
method~\cite{pseudo} with $N^2$ collocation points and the $2/3$ dealiasing
rule. Here $\psi$ is the stream function, $\omega$ the vorticity,
$J(\psi,\omega) \equiv (\partial_x \psi)(\partial_y \omega) - (\partial_x
\omega) (\partial_y \psi)$, $\nu$ the kinematic viscosity, and $\mu$ the
coefficient of Ekman friction. At the point ($x$,$y$) the velocity $\bfu
\equiv (-\partial_y \psi, \partial_x \psi) $ and the vorticity $\omega = \lap
\psi$; the external deterministic force $f_\omega(x,y)=-F_0\kinj\cos(\kinj
x)$, where $F_0$ is an amplitude and $\kinj$ the energy-injection scale in
Fourier space.  The injected energy displays an inverse cascade to small $k$.
Ekman friction removes energy from all Fourier modes; in particular, it
removes energy from small-$k$ Fourier modes in such a way that the system
reaches a nonequilibrium statistically steady state.  We evolve
Eq.~\ref{eqch6:nsvor1} in time by a second-order, exponential Runge-Kutta
method~\cite{cox+mat02}.  In all our simulations we wait for a time $\Ttran$
[see the caption of Table~(\ref{tablech6:para})] to allow transients to die 
out so that our system reaches a statistically steady state.
Figure~(\ref{fig:snap}) shows a typical pseudocolor plot of $\psi$ in such 
a state.

To calculate Lagrangian quantities we track $\np$ particles. The
evolution equation for a Lagrangian particle is
$  \frac{d}{dt}{\bf x}_L(t) = {\bf u}({\bf x}_L,t), $
where ${\bf x}_L(t)$ is the position of a Lagrangian particle at time $t$;
${\bf u}({\bf x_L},t)$, the velocity at the Lagrangian particle position, is
evaluated from the Eulerian velocity field ${\bf u}({\bf x},t)$ by using a
bilinear-interpolation scheme~\cite{nr}.  
The evolution equation of the
particles is solved by a second-order, Runge-Kutta method~\cite{nr}.
Initially all the particles are seeded randomly into the flow. A typical
Lagrangian-particle track superimposed on a representative pseudocolor plot
of $\psi$ is shown in Fig.~(\ref{fig:snap}). The list of
parameters used in our simulations is given in Table~(\ref{tablech6:para}).

From the velocity-gradient tensor $\mathcal{A}$, with components $A_{ij}
\equiv \partial_i u_j$, we obtain the Okubo-Weiss parameter $\Lambda$, the
discriminant of the characteristic equation for $\mathcal{A}$. 
If $\Lambda$ is
positive~(negative) then the flow is
vortical~(extensional)~\cite{oku70+wei91}. In an incompressible flow in two
dimensions $\Lambda = {\rm det} \mathcal{A}$; 
and the PDF of $\Lambda$ has been
shown~\cite{per+pan09} to be asymmetrical about $\Lambda=0$ (vortical regions
are more likely to occur than strain-dominated ones). For the Eulerian case,
we monitor the time evolution of $\Lambda$ at $\np$ randomly chosen points
that are fixed in the simulation domain. In our Lagrangian study, we begin
with the values of $\Lambda$ on the spatial grid that we use for our Eulerian
DNS; bilinear interpolation then yields $\Lambda$ at the positions of
Lagrangian particles, which can be at points that do not lie on this grid; we
can thus monitor the evolution of $\Lambda$ along Lagrangian-particle
trajectories.

We denote the presistence-time PDFs by $P$ and the associated cumulative PDFs
by $Q$; the subscripts $E$ and $L$ on these PDFs signify Eulerian and
Lagrangian frames, respectively; and the superscripts $+$ or $-$ distinguish
PDFs from vortical points from those from extensional ones.  To find out the
persistence-time PDF $P_E^+(\tau)$ [resp., $P_E^-(\tau)$] we analyse the
time-series of $\Lambda$ obtained from each of the $\np$ Eulerian points and
construct the PDF of the time-intervals $\tau$ over which $\Lambda$ remains
positive (resp., negative).  The same method applied to the time series of
$\Lambda$, obtained from each of the $\np$ Lagrangian particles, yields
$P_L^+(\tau)$ [resp., $P_L^-(\tau)$].

\begin{figure*}
\begin{center}
\includegraphics[width=0.3\linewidth]{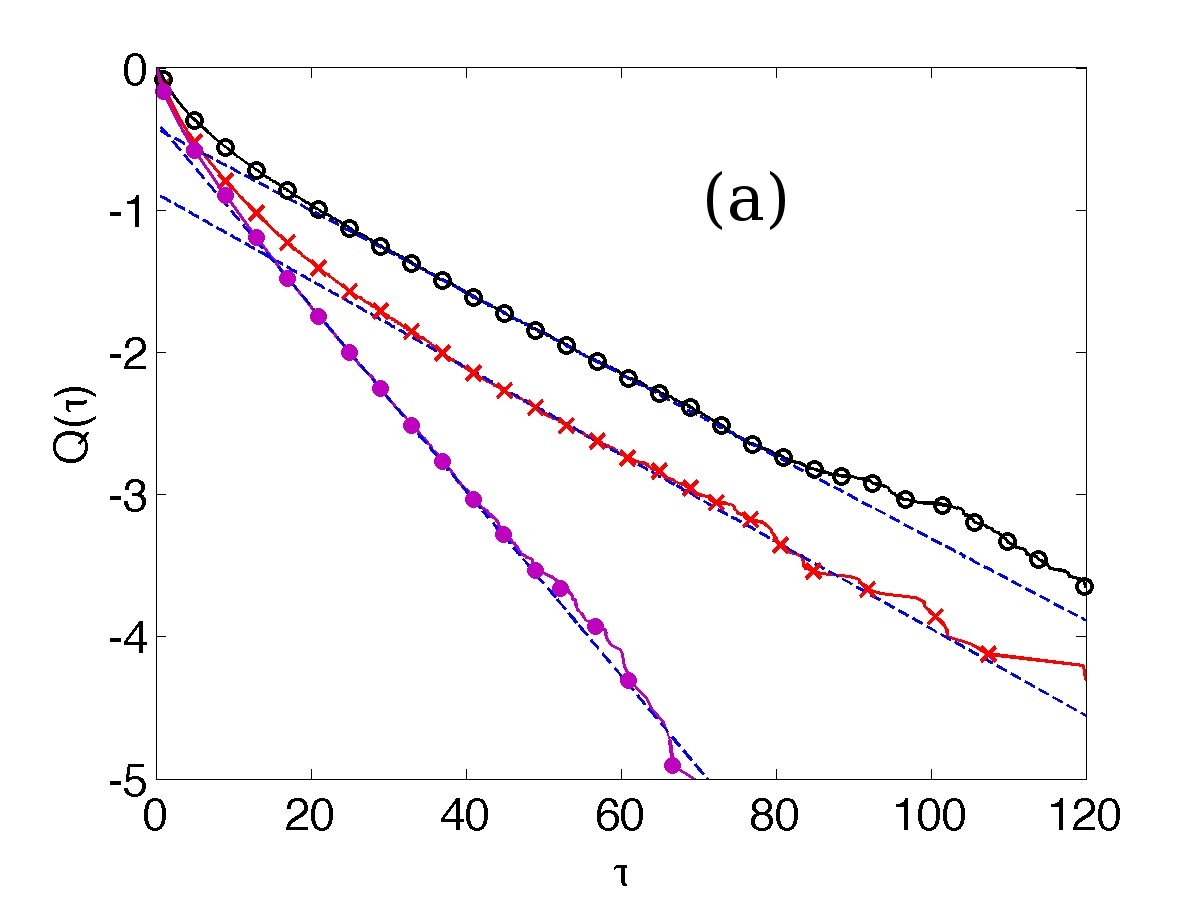}
\includegraphics[width=0.3\linewidth]{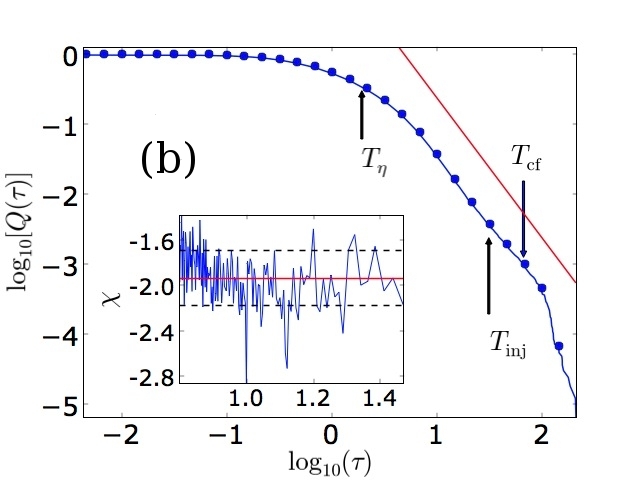}
\includegraphics[width=0.3\linewidth]{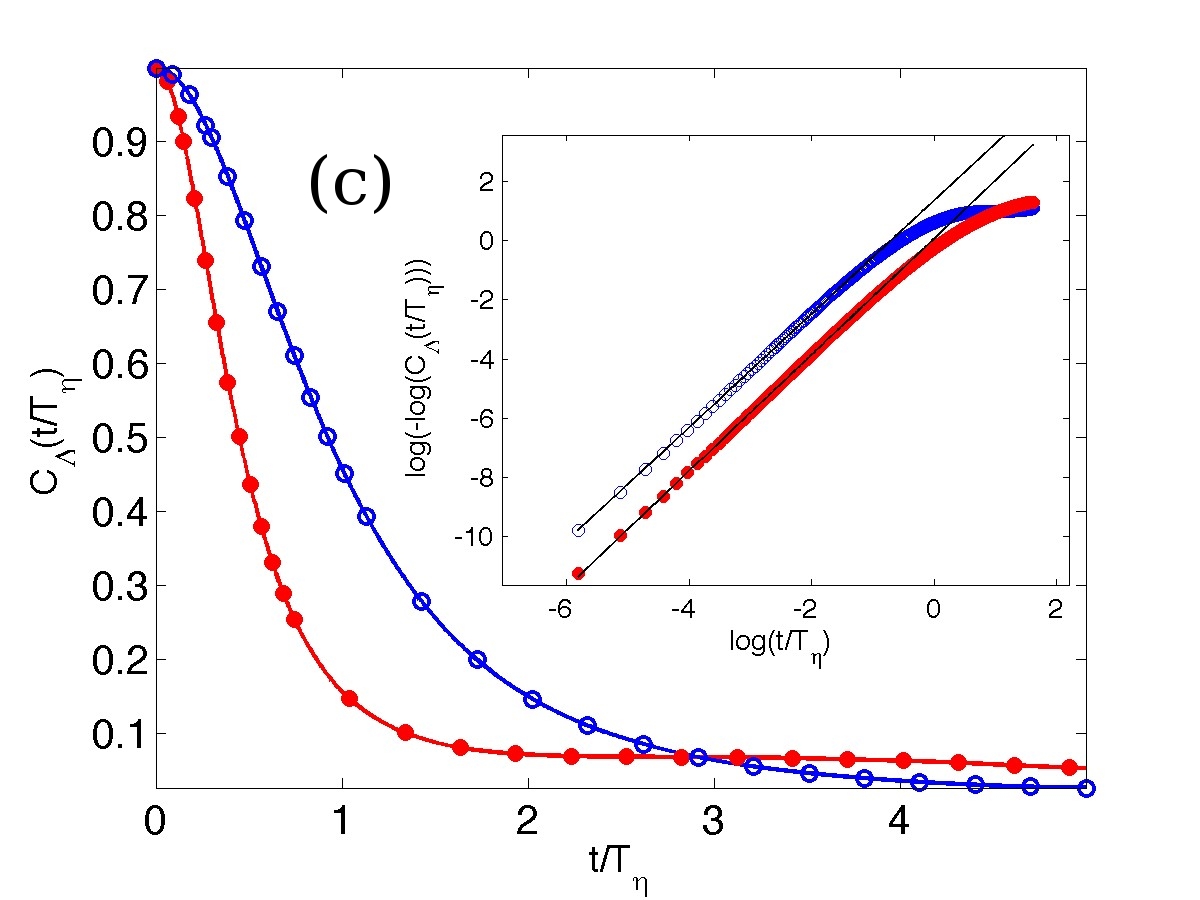}
\caption{\label{fig:three} (Color online) 
(a) Representative semilog plots
(base 10) of the cumulative persistence-time PDFs $\qpe$ (red crosses),
$\qne$ (black open circles), and $\qnl$ (magenta full circles); the
subscripts $E$ and $L$ signify Eulerian and Lagrangian frames, respectively;
and the superscripts $+$ or $-$ distinguish PDFs from vortical points from
those from extensional ones.  The dashed lines indicate exponential fits to
these cumulative PDFs.  These plots use data from our run {\tt R4}.
(b) A representative log-log plot 
(base 10) of the cumulative PDF $\qpl(\tau)$(\textbullet) versus $\tau$ for 
our run ${\tt R4}$; the full red line, with a slope equal to $-2$, is 
drawn for reference. The vertical arrows indicate the time scales
(from left to right) $\Teta$, $\Tinj$, and $\Tcutoff$, respectively.
The inset shows the local slope $\chi=d\log_{10}
\qpl(\tau)/d\log_{10}(\tau)$ versus $\tau$; the horizontal line is drawn at  
the mean  $\langle \chi \rangle\simeq -1.93$ and the black dashed lines,  
drawn at  $\langle \chi \rangle \pm \sigma_\chi$, where 
$\sigma_\chi \simeq 0.2$ is the standard deviation of $\chi$.
(c) Plots of the autocorrelation function of the 
Okubo-Weiss parameter $C_\Lambda(t/\Teta)$ (see text) versus $t/\Teta$ 
in Eulerian (blue open circles) and Lagrangian (red full circles) frames 
for our run ${\tt R1}$. In the inset we compare the data points for these
plots with their fits (full lines) to the form $G \exp[-(t/T_{\Lambda})^2]$. }
\end{center}
\end{figure*}

We use the rank-order method~\cite{mit05a} to calculate cumulative PDFs 
because they are free from binning errors. 
We show representative
plots of the cumulative PDFs $\qpe$ (red crosses), $\qne$ (black open
circles), and $\qnl$ (magenta full circles) in Fig.~(\ref{fig:three}a); the
dashed lines indicate exponential fits to these cumulative PDFs.  From these
and similar fits we conclude that the PDFs $\ppe$, $\pne$, and $\pnl$ have
exponentially decaying tails, for all the runs, from which we can extract
characteristic time scales. In particular, from the Eulerian PDFs we obtain
the times $\TpE$ and $\TmE$, for vortical and strain-dominated regions,
respectively, which are defined as follows: $\ppe(\tau) \sim
\exp(-\tau/\TpE)$ and $\pne(\tau) \sim \exp(-\tau/\TmE)$ as $\tau \to
\infty$ [see Table ~(\ref{tablech6:para})].
The Lagrangian PDF $\pnl$ also has an exponentially decaying
tails from which we obtain the characteristic time $\TmL$
[see Table ~(\ref{tablech6:para})].

The persistence-time PDF $\ppl$ and the associated cumulative PDF $Q_L^+$ of
a Lagrangian particle in a vortical region is very different from those
discussed above: the tails of $\ppl$ and $Q_L^+$ have power-law, and not
exponential, forms; we show this in Fig.~(\ref{fig:three}b) via a
representative plot $Q_L^+$ for the run ${\tt R4}$. Thus, as in
nonequilibrium statistical mechanics~\cite{maj99}, we can define the
persistence exponent $\theta$ via $\ppl \sim \tau^{-\theta}$.  
Our run ${\tt R4}$ has the largest value of $Re_\lambda$ amongst the runs
${\tt R1-4}$ [Table~(\ref{tablech6:para})] and, therefore, is best suited for
estimating $\theta$ from plots such as the one in Fig.~(\ref{fig:three}b).
We obtain the exponent $(-\theta+1)$ by fitting a power-law to the tail 
of the cumulative PDF $Q_L^+$. To find the best estimate for $(-\theta+1)$
we evaluate the local slope $\chi=d\log_{10}Q_L^+(\tau)/d\log_{10}(\tau)$ in 
the region shown in the inset of Fig.~(\ref{fig:three}b). Our estimate 
for $(-\theta+1)$ is the mean value of $\chi$ over the region indicated 
in the inset; the standard deviation of $\chi$ yields the error; finally we 
obtain $\theta = 2.9\pm 0.2$.
Our other runs ${\tt R1-3}$ yield smaller 
scaling ranges than the one in Fig.~(\ref{fig:three}b) and, therefore,
yield values for $\theta$ with larger error bars; but these
values are consistent with our estimate for $\theta$ from run ${\tt R4}$.  
 We also find that the persistence exponent $\theta$ does not depend
on the parameters $\mu$, $F_0$, and $\kinj$ (within error-bars) in the
range of parameters accessible in our simulations.  Based on this evidence we
conjecture that $\theta$ is a new universal exponent that characterizes
two-dimensional, Navier--Stokes turbulence. A conclusive proof of this
conjecture must await future studies.

The exponent $\theta>1$, so we can obtain the average lifetime
$\Tmean$ of a particle in a vortical region from $\ppl$; 
Another estimate
of the lifetime of vortices in the Lagrangian frame is given by
the time scale $\Tcutoff$, which is the cutoff scale of the
power-law decay of tail of $\ppl$,
[Table~(\ref{tablech6:para})]. 

In most persistence problems in nonequilibrium statistical mechanics the
power-law tail for the presistence-time PDF appears in the following way:
Typically we consider the persistence-time PDF of $\phi$ that comes from a
Gaussian, but {\it non-stationary}, process; a transformation to logarithmic
time $s$ transforms such a process to a Gaussian stationary process (GSP)
$X(s)$; if, in addition, the GSP is also Markovian, i.e., the autocorrelation
function of the GSP, $f(t) = \bra{X(s)X(s+t)}_s$, is an exponential function
of $t$, then the presistence-time PDF of the GSP can be shown to have an
exponential tail~\cite{maj99}. If we now transform back from logarithmic to
linear time, the exponential tail is transformed to a power-law tail.
Furthermore, for a GSP for which the autocorrelation function $f(t)$ decays
faster than $1/t$ for large $t$, the persistence-time PDF $P(t) \sim
\exp(-\gamma t)$, where $\gamma$ is a constant~\cite{maj99}.  

In the two-dimensional, fluid-turbulence problem that we study here, we have
checked numerically that $\Lambda$ is a stationary process; hence we do not
need to transform to logarithmic time.  We have calculated two types of
autocorrelation functions for $\Lambda$ (we denote these generically by
$C_\Lambda(t)$ in  Fig.~(\ref{fig:three}c)): (a) For the first we evaluate
$\bra{\Lambda(0)\Lambda(t)}$ over the track of a Lagrangian particle; this is
shown by blue open circles in Fig.~(\ref{fig:three}c); (b) for the second we
evaluate $\bra{\Lambda(0)\Lambda(t)}$ at a given point on our Eulerian grid;
this is shown by red full circles in Fig.~(\ref{fig:three}c); here
$\bra{\cdot}$ denotes averages over different origins of time and also over
$\np$ different Lagrangian particles, for case (a), or over $\np$ different
Eulerian positions, for case (b).  As we show in the inset of
Fig.~(\ref{fig:three}c) for both these cases, $C_\Lambda(t)$ is approximated
well by the function $\exp[-(t/T_{\Lambda})^2]$, over the range  $10^{-4} <
(t/\Teta) < 10^{-1}$; this decay is clearly faster than $1/t$ for large $t$.
Here $T_{\Lambda}$, the characteristic decay time, is slightly larger in the
Lagrangian case than in the Eulerian one; however, in both these cases
$T_{\Lambda}\simeq\Teta$.  Plots of $C_{\Lambda}$ from our other runs ${\tt
R1-3}$ are similar to the one in Fig.~(\ref{fig:three}c) from run ${\tt R4}$.
Note that, in the problem we study, the persistence-time PDF is not
constrained to have an exponential tail because $\Lambda$, although
stationary, is not a Gaussian process.  

We have presented a natural framework for studying the persistence problem in
two-dimensional fluid turbulence. The most important result of our study is
that the persistence-time PDFs for vortical points show qualitatively
different behaviors in the Eulerian and Lagrangian cases: In vortical
regions, for the Eulerian case, this PDF displays an exponential decay; in
contrast, for the Lagrangian case it shows a power-law tail. Qualitatively
such nontrivial behavior appears because a passive particle can be trapped
for quite some time in a vortical 
region~\cite{movie})
Furthermore, we provide a way of
measuring the lifetime of a vortex precisely. In the Eulerian frame the
characteristic lifetime of a vortex is the time scale $\TpE$ that follows
from the exponential form $\ppe(\tau) \sim \exp(-\tau/\TpE)$.  However, in
the Lagrangian frame the persistence-time PDF for vortical points shows a
power-law decay with a persistence-time exponent 
$\theta =2.9\pm 0.2$.  
Hence
there is no single time scale which describes the time spent by passive
particles in vortical points; but, as we have mentioned above, an average
residence time can be defined because $\theta> 1$.  

The PDF of residence times of passive tracers in vortical regions $\ppl$ is of
great fundamental and engineering importance.  Earlier
studies~\cite{car96,bey01} have attempted to measure this PDF; however, their
methods of obtaining it are not as precise as the one we present here: the
Okubo-Weiss parameter $\Lambda$, which we employ, helps us to distinguish
clearly between vortical and extensional regions in the two-dimensional flows
we consider.
The natural way of generalizing our study 
to its three-dimensional counterpart is to replace
the Okubo-Weiss parameter by $QR$ plots~\cite{cantwell93} and then to study
the PDFs of residence times of Lagrangian particles in each quadrant of the
$QR$ plot; here $Q = -\frac{1}{2}tr(\mathcal{A}^2)$ and $R =
-\frac{1}{3}tr(\mathcal{A}^3)$ 
where $\mathcal{A}$ is the velocity gradient matrix. 
It remains to be seen
whether the tails of such PDFs have power-law tails in three-dimensional
turbulence; such a study lies beyond the scope of this paper.
From the point of view of the general theory of persistence problems, the
time-series of $\Lambda$ in the Lagrangian frame provides a particularly
interesting example. The persistence-time PDF has a power-law behavior for
positive $\Lambda$ but an exponential tail for negative $\Lambda$.  A
similar, but less dramatic, example of such asymmetry has been observed for
the case of growing interfaces whose height $h$ obeys the Kardar-Parisi-Zhang
(KPZ) equation~\cite{kal+kru99}.  For this KPZ case, the persistence-time PDF
shows power-law tails, but with different persistence-time exponents for
positive and negative $h$. 
We hope our study will stimulate new experimental investigations of
persistence-time PDFs in two-dimensional fluid turbulence and also studies of
such PDFs for other non-Gaussian, stationary processes.

We thank S. Krishnamurthy and D. Sanyal for discussions, 
the European Research Council under the AstroDyn Research Project No.\ 227952,
CSIR, UGC, and DST (India) for support
and SERC (IISc) for computational resources. PP and RP are members
of the International Collaboration for Turbulence Research; RP, PP,
and SSR acknowledge support from the COST Action MP0806.
\begin{table*}
\begin{center}
\begin{tabular}{@{\extracolsep{\fill}} c c c c c c c c c c c c c c c c c c} 
\hline 
$Run $ &$N$ & $\nu$ & $\mu$ & $F_0$ & $\kinj$ & $l_d$ & $\lambda$ &
  $Re_{\lambda}$  & $\Teddy$ & $\Teta $ & $\TmE$ & $\TmL$ & 
     $\TpE$ & $\Tmean$ &  $\Tinj$ & $\Tcutoff$  \\
\hline \hline
${\tt R1}$ & $512$ & $0.016$ & $0.1$ & $45$ &$10$ &
$2.3\times10^{-2}$ & $0.2$  & $59$  & $0.1$ & $3.4\times10^{-2}$
& $0.3\pm0.04$ & $0.12\pm0.02$ & $0.21\pm0.05$ & $2.8\times10^{-2}$ & 
$0.3$ & $0.9$ \\ 
${\tt R2}$ & $512$ & $0.016$ & $0.45$ & $45$ &$10$ & 
$2.1\times10^{-2}$ & $0.1$  & $27$  & $0.1$ & $2.7\times10^{-2}$ 
& $0.4\pm0.05$ & $0.17\pm0.02$ & $0.24\pm0.03$ & $4.2\times10^{-2}$ & 
$0.2$ & $0.8$ \\ 
${\tt R3}$ & $1024$ & $10^{-5}$ & $0.01$ & $0.005$ & $10$ & 
$4.3\times10^{-3}$ & $0.1$  & $827$  & $11$ &  $1.9$ 
& $19\pm3$ & $10\pm2$ & $13\pm2$ & $1.8$ & 
$19.9$ & $76$ \\
${\tt R4}$ & $1024$ & $10^{-5}$ & $0.01$ & $0.005$ & $4$ & 
$5.4\times10^{-3}$ & $0.2$  & $1319$  & $7$ &  $2.9$ 
& $31\pm5$ & $15\pm2$ & $25\pm4$ & $2.5$ & 
$30.2$ & $81$ \\
\hline 
\end{tabular}
\end{center}
\caption{ 
Parameters for our runs {\tt R1-4}: $N$ is the number of grid points 
along each direction, $\np=1000$ is the number of Lagrangian particles and 
Eulerian positions (at which we monitor $\Lambda$), $\nu$ the 
kinematic viscosity, $\mu$ the Ekman friction, $F_0$ the forcing amplitude, 
$\kinj$ the forcing wavenumber, $l_{\rm d} \equiv (\nu^3/\varepsilon)^{1/4}$ 
the dissipation scale, $\lambda\equiv \sqrt{\nu E/\varepsilon}$ the 
Taylor microscale, $Re_{\lambda}\equiv \urms \lambda/\nu$ the 
Taylor-microscale Reynolds number, 
$\Teddy\equiv[\pi \sum_k(E(k)/k)/(2 \urms^2)]$ 
the eddy-turn-over time, and
$\Teta \equiv \sqrt{\nu/\varepsilon}$ the Kolmogorov time scale.
The time scales $\TmE$, $\TmL$ and 
$\TpE$ are obtained from exponential fits to the tails of the cumulative
PDFs $\qne$, $\qnl$ and $\qpe$, respectively, as shown in 
Fig~(\ref{fig:three}a). 
$\Tmean$ is the average time spent by a Lagrangian particle in a 
vortical region, $T_{\rm inj}\equiv (l_{\rm inj}^2/E_{\rm inj})^{1/3}$ is the 
energy-injection time scale, where $E_{\rm inj}=<{\bf f_{\rm u}}\cdot {\bf u}>$,
(${\bf f}_{\omega} = \curl {\bf f}_{\rm u}$),
 is the energy-injection rate and $l_{\rm inj}=2\pi/\kinj$ is the 
energy-injection length scale.  $\Tcutoff$ is the large-time cutoff of the
scaling range of $\qpl$ as shown in Fig.~\ref{fig:three}b.
We do not use data from the initial period of duration $\Ttran=100\Teddy$; 
this removes the effects of transients. We use a square simulation domain 
with side $L=2\pi$ and grid spacing $\delta_x = L/N$.
} 
\label{tablech6:para}
\end{table*}
\end{document}